\documentstyle[aps,prl,epsf,epsfig,floats]{revtex}

\begin{document}

\draft

%%%%%%%%%%%%%%%%%%%%%%%%%%%%%%%%%%%%%%%%%%%%%%%%%%%%%%%%%%%
%                                                         %
%   Comment out \wideabs{ when not using twocolumn mode   %
%                                                         %
%%%%%%%%%%%%%%%%%%%%%%%%%%%%%%%%%%%%%%%%%%%%%%%%%%%%%%%%%%

\wideabs{
\title{Dipole relaxation losses in DNA}

\author{M. Briman, N.P. Armitage, E. Helgren, and G. Gr\"{u}ner}

\address{ Department of Physics and Astronomy,
University of California, Los Angeles, CA 90095}
\date{\today}
\maketitle
\begin{abstract}

The electrodynamic response of DNA in the millimeter wave range is
investigated. By performing measurements under a wide range of
humidity conditions and comparing the response of single strand DNA and double strand DNA, we show that the appreciable AC
conductivity of DNA is not due to photon activated hopping between
localized states, but instead due to dissipation from dipole
motion in the surrounding water helix. Such a result, where the
conductivity is due to the constrained motion of overdamped
dipoles, reconciles the vanishing DC conductivity of DNA with the
considerable AC response.

\end{abstract}


} % end of \wideabs

The electrical conductivity of DNA has been a topic of much recent
interest and controversy \cite{Dekker}. Measurements from
different groups have reached a variety of conclusions about the
nature of charge transport along the double helix.  DNA has been
reported to be metallic \cite{Fink}, semiconducting \cite{Porath},
insulating \cite{Braun,dePablo}, and even a proximity effect
induced superconductor\cite{Kasumov}. However, questions have been
raised with regards to the role played by electrical contacts,
length effects, and the manner in which electrostatic damage,
residual salt concentrations, and other contaminations may have
affected these results \cite{Dekker}.  More recent measurements,
where care was taken to both establish a direct chemical bond
between $\lambda$-DNA and Au electrodes and also control the
excess ion concentration, have given compelling evidence that the
DC resistivity of the DNA double helix over long length scales
($<10\mu$m) is very high indeed ($\rho > 10^6 \Omega - cm$)
\cite{Zhang}.  Such DC measurements contrast with recent
contactless AC measurements that have shown that there is
appreciable conductivity at microwave and far-infrared frequencies
\cite{Tran,Helgren} the magnitude of which approaches that of a
well-doped semiconductor \cite{HelgrenSiP}.

Previously, the AC conductivity in DNA was found to be well
parameterized as a power-law in $\omega$ \cite{Tran,Helgren}. Such
a dependence can be a general hallmark of AC conductivity in
disordered systems with photon assisted hopping between random
localized states \cite{ES} and led to the reasonable
interpretation that intrinsic disorder, counterion fluctuations,
and possibly other sources created a small number of electronic
states on the base pair sequences in which charge conduction could
occur. However, such a scenario would lead to thermally activated
hopping conduction between localized states and is thus
inconsistent with the very low DC conductivity\cite{Zhang}.  A
number of outstanding issues arise:  Are there localized regions
along the helix where a continuous conducting path is not present,
but still AC hopping between localized states over distances of a
few base pairs can occur?  Are there sensitive length dependencies
in the DNA strands?  Is there a difference between between the
samples of various groups?  Perhaps different charge conduction
mechanisms play a role at finite frequency.

To the end of resolving some of these matters, we have performed
AC conductivity experiments in the millimeter wave range under a
wide range of humidity conditions.   We show that the appreciable
AC conductivity of DNA in the microwave and far infrared regime
should not be viewed as some sort of hopping between localized
states and is instead likely due to dissipation in the dipole
response of the water molecules in the surrounding hydration
layer.  It can be well described by a Debye-like relaxation of
water molecules in the surrounding water helix. At low humidities
the response is well modelled by considering the rotation of
single water molecules in the structural water layer. As the
number of water molecules per base pair increases, dissipation due
to the collective motion of water dipoles increases, until
eventually the conductivity resembles that of bulk water.  By
measuring both single strand (ssDNA) and double strand DNA (dsDNA)
over a wide range of humidities we are able to show that, at least
in principle, all the AC conductivity of DNA can be assigned to
relaxation losses of water dipoles. This result reconciles the
apparent complete lack of DC conductivity with the appreciable AC
response.

Double stranded DNA films were obtained by vacuum drying of 7mM
PBS solution containing 20 mg/ml sodium salt DNA extracted from
calf thymus and salmon testes (Sigma D1501 and D1626).  The
results were found to be independent of the use of calf or salmon
DNA. Our choice for these concentrations deserves further
explanation. It is well known that at a given temperature double
helical conformation of DNA can exist in solution only with a
certain concentration of positive ions. Excess salt cannot be
removed by vacuum drying, so large amounts of residual salt in
films could introduce significant errors in conductivity, due to
both the ionic conduction of the salt itself and its additional
hydration during humidity changes. Melting temperature
calculations \cite{melt1,melt2} for long native pieces of DNA with
C-G content around 40\% show that 2-10 mM concentration of sodium
cations is enough to stabilize the double helix at room
temperature.  Films were prepared with differing salt amounts and
it was found that as long as the excess salt mass fraction is kept
between 2-5\% the final results were not significantly affected.
In order to improve the DNA/salt mass ratio we used a high
concentration of DNA, but 20 mg/ml appears to be the limit. Higher
concentrations makes it difficult for DNA fibers to dissolve and
the solution becomes too viscous, which prevents producing the
flat uniform films which are of paramount importance for the
quasi-optical resonant technique. Single stranded DNA films were
prepared from the same original solution as the double stranded
ones, with preliminary heating up to 95 C for 30 minutes and fast
cooling down to 4 C. In both dsDNA and ssDNA cases the
conformational state was checked by fluorescent microscope
measurements. The dry films were 20 to 30 microns thick and were
made on top of 1mm thick sapphire windows. Immediately after
solution deposition onto the sapphire substrates the air inside
the viscous solution was expelled by vacuum centrifuging at 500g,
otherwise the evaporation process causes the formation of air
bubbles that destroy the film uniformity.

The AC conductivity was measured in the millimeter spectral range.
Backward wave oscillators (BWO) in a quasi-optical setup (100 Ghz
- 1 THz) were employed as coherent sources in a transmission
configuration.  This difficult to access frequency range is
particularly relevant as it corresponds to the approximate
expected time frame for relaxation processes in room temperature liquids (1-10 ps).
Importantly, it is also below the energy range where one expects
to have appreciable structural excitations.  The technique and
analysis are well established \cite{Schwartz}.  We utilize the
fact that for plane waves incident normally on a slab of material,
transmission resonances occur when the slab is an integer number
of half wavelengths. Thus, using a $\approx 1$ mm sapphire disc as
a substrate, resonances occurred approximately every 50 GHz.
Having analyzed the transmission through the sapphire alone prior
to mounting the sample, the optical properties of the substrate
were well characterized.  Thus using a two-layer transmission
model, each resonance can be analyzed to extract the optical
properties of the DNA film, allowing for a 1.5 $cm^{-1}$
resolution of the spectra.

Samples were measured at room temperature at several fixed
humidity levels which were maintained by putting them in a hermetically sealed environment with a saturated salt solution \cite{Falk}.
The change in thickness and mass of the DNA films at different
humidities were tracked by separate measurements within a
controlled environment for each sample in a glove box.  The total
number of water molecules per nucleotide $A$ can be correlated to
the relative humidity x (x=0-1) through the so-called
Branauer-Emmett-Teller (BET) equation \cite{BET}

\begin{equation}
A=\frac{BCx}{(1-x)(1-x+Cx)}.
\end{equation}

The constant $B$ is the maximum number of water molecules in the
first layer sites.  According to the statistical formulation of
the BET equation by Hill \cite{Hill}, mobile water molecules
within the double helix can be characterized as 2 types. The first
are ones within the initial hydration layer, which are directly
attached to DNA and have a characteristic binding energy
$\epsilon_1$. Water molecules of the second and all other layers
can be approximated as having a binding energy $\epsilon_L$.  To a
good approximation this $\epsilon_L$ can be taken to be that of
bulk water.  These parameters enter into the BET equation through
the expression for $C$ which equals
$De^{(\frac{\epsilon_1-\epsilon_L}{kT})}$ where $D$ is related to
the partition function of water.  Also we should note that there
is, in actuality, a structural 0-th layer of water molecules,
containing 2.5-3 water molecules per nucleotide that cannot be
removed from the helix under typical conditions ~\cite{Tao}.

\begin{figure}[htb]
\centerline{\epsfig{figure=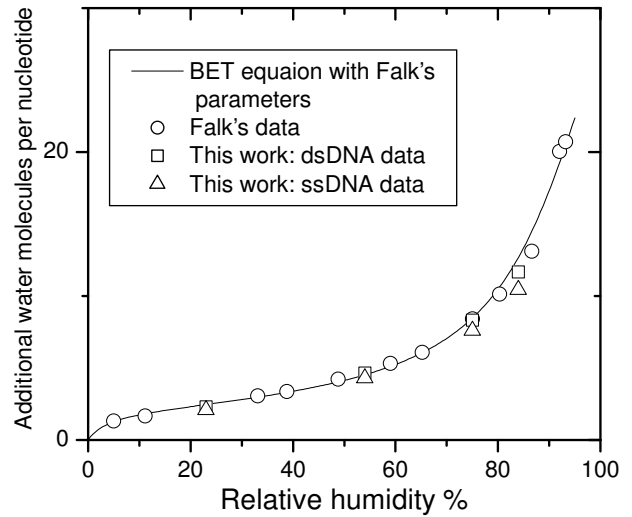,width=9cm}} \vspace{.2cm}
\caption{Absorbtion of water molecules per nucleotide as a
function of humidity.  The data represented by the open circles is taken from Falk \textit{et. al.}} \label{hydration}
\end{figure}

That it is reasonable that the mobile water layers of DNA can be modelled
by distinguishing 2 different sets of water parameters was first
established by Falk \textit{et al.}'s ~\cite{Falk} use of the BET
equation to describe the hydration of sodium and lithium DNA salts
from calf thymus and salmon testes. They found good agreement
between experimental data and theory with constants $B=2.2$ and
$C=20$. We performed a similar hydration study of our dsDNA and
ssDNA films; as shown in Fig. \ref{hydration} the hydration of our
films are perfectly consistent with Falk's result.  Note that
there is no appreciable difference in the hydration between dsDNA
and ssDNA.

\begin{figure}[htb]
\centerline{\epsfig{figure=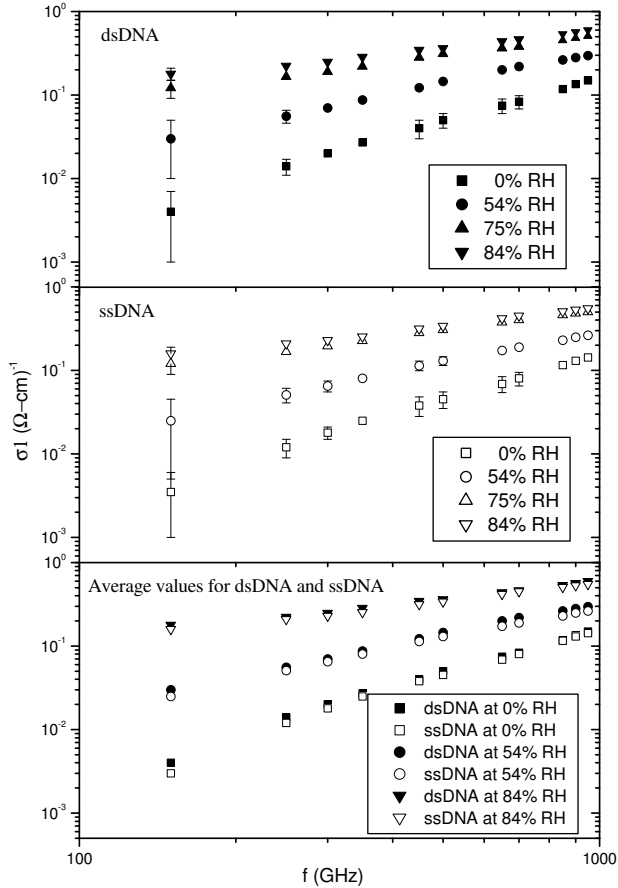,width=9cm}} \vspace{.2cm}
\caption{Frequency dependence of the conductivity of calf thymus
DNA at different relative humidity levels.  (a)  Double stranded
DNA (b)  Single stranded DNA  (c) A comparison of conductivity
between single and double stranded DNA.} \label{conduct}
\end{figure}

In Fig. \ref{conduct} we present data for the extracted
$\sigma_1(\omega)$ of both dsDNA and ssDNA thin films. One can see
that in both cases, the conductivity is an increasing function of
frequency.  Since the conductivity is also an increasing function
of humidity, one may wish to try to seperate the relative
contributions of charge motion along the DNA backbone from
that of the surrounding water molecules.

First, one can consider that there should be two main effects of
hydration in our dsDNA films. There is the hydration itself, where
water molecules are added in layers to the double helix; this is
well described by BET equation \cite{BET}. Additionally, the
conformational state of dsDNA also changes as a function of
humidity. For example, sodium salt calf thymus DNA is in a B-like
disordered form at humidities from 0-40\%, above which it
transfers to the A form, and finally to a well ordered B-form at
humidities higher than 80\% \cite{Maleev,Lindsay}. Additional
water molecules certainly contribute to the increase in
conductivity, but at high humidities there is the possibility that
some of the conduction might be due to an increase in electron
transfer along the dsDNA helix in the ordered B form.  However
since such an effect would be much reduced in disordered and
denaturalized ssDNA films and since Fig. 2 shows that to within
the experimental uncertainty the conductivity of dsDNA and ssDNA
in the millimeter wave range is identical, it is most natural to
suggest that water is the major contribution to the AC
conductivity.  From this comparison of dsDNA and ssDNA, we find no evidence for charge conduction along the DNA backbone.

In Fig. 3 we plot the the conductivity $\sigma_1$ of the DNA films
normalized by the expected volume fraction of water molecules
including both the hydration layers plus the structural water.
Although this normalization reduces the spread in the thin film
conductivity at the lowest frequencies it does not reduce it to
zero, showing that if the largest contribution to the conductivity
comes from water, the character of its contribution changes as a
function of humidity.

The complex dielectric constant of bulk water has been shown to be
well described by a biexponential Debye relaxation model
\cite{Ronne,Kindt,Barthel}, where the first relaxation process
\cite{Ronne}, characterized by a time scale $\tau_D=8.5$ ps,
corresponds to the collective motion of tetrahedral water
clusters, and the second from faster single molecular rotations
\cite{Agmon} with a time scale $\tau_F=170$ fs. For bulk water,
the contribution of each relaxation process is determined by the
static dielectric constant $\epsilon_S(T)=87.91e
^{-0.0046T[^oC]}$, $ \epsilon_1=5.2$, and the dielectric constant at
high frequencies $\epsilon_{\infty}=3.3$.

\begin{equation}
\widehat{\epsilon}(w)=\epsilon_{\infty}+\frac{\epsilon_S-\epsilon_1}{1+i\omega\tau_D}+
\frac{\epsilon_1-\epsilon_{\infty}}{1+i\omega\tau_F}
\end{equation}

When applying Eq. 2 to the dipole relaxation losses of DNA,
one expects that the relative contributions of the two frequency dependent terms will
change as increasing humidity increases the average effective
coordinate number.  For instance, at 0\% humidity it is reasonable
to assume that the first term which is due to the collective
motion of water clusters, cannot play a role as the structural
water is not tetrahedrally coordinated.  For high hydration
levels, where multiple water layers exist around the dipole helix,
the relaxation losses of the water layer may approach those of
bulk water.  We can compare the above equation using the
independently known values \cite{Ronne} for $\tau_D$,
$\epsilon_S$, $\tau_F$ and $\epsilon_1$ to the experimental data
normalized to the expected volume fraction of the water from the
independently determined water uptake curves shown in Fig. 1.  In
Fig. 3, along with the experimental data at two representative
humidity levels, two theoretical curves for 0\% and 100\% humidity
are plotted. With the only two assumptions being that at 0\%
humidity, the sole relaxational losses come from singly
coordinated water molecules in the structural water layer and that
it is only at higher humidity levels where the collective losses
can gradually play a greater role, the theoretical curves provide a very good fit to the data over almost all of the
measured frequency range. At low humidity the data is well matched
by the theory incorporating only single molecule rotations.  At
high humidity, the data begins to approach the behavior of 'free'
water.  For these two limits the theoretical curves have no free
parameters.

\begin{figure}[htb]
\centerline{\epsfig{figure=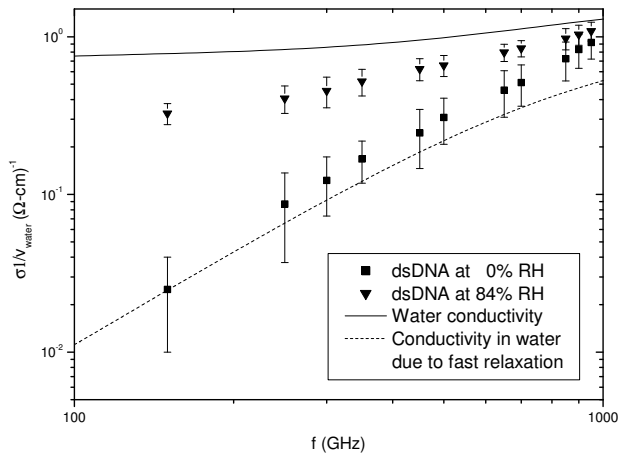,width=9cm}} \vspace{.2cm}
\caption{Conductivity of dsDNA and ssDNA films normalized by the
volume fraction of all water molecules (structural plus hydration
layer). For clarity, only 0\% and 84\% humidities are shown.  The
solid line represents the conductivity of pure water as modelled
by the biexponential Debye model using the parameters of Ronne et
al. The dashed line shows just the contribution from single water
molecule relaxation.} \label{adjcond}
\end{figure}

The only appreciable discrepancy between theory and experiment is
the high frequency data at low humidity, where the biexponential
Debye model underestimates the conductivity. This may be due to
a number of reasons.  At very low relative humidities it is
possible for the ionic phosphate groups on the DNA backbone to form stable
dihydrates which may give their own contribution to relaxation
losses through their additional degree of freedom \cite{Falk}.
Alternatively, it may also be that at higher frequencies for low
hydration samples, the weak restoring force from charge-dipole
interaction in the structural water layer begins to become more
significant and our biexponential Debye model becomes less
applicable.

In conclusion, we have found that the considerable AC conductivity
of DNA can be largely ascribed to relaxational losses of the
surrounding water dipoles.  The conductivity of ssDNA and dsDNA
was found to be identical to within the experimental error,
indicating that there is essentially no charge conduction along
the DNA backbone itself.  The conclusion that the observed
conductivity derives from the water layer is supported by the
fact that, over much of the range, it can be well described by a
biexponential Debye model, where the only free parameter is the
relative contributions of single water molecule and tetrahedral water cluster
relaxation modes.  Generally speaking, because many large biomolecules have surrounding water layers, a result such as ours shows that one must be aware of the possibility of such relaxation losses when investigating the electrodynamic response of such systems.

We would like to thank K. Greskoviak for help with sample
preparation.  The research at UCLA was supported by the National
Science Foundation grant DMR-0077251.

\end{document}